\documentclass[pre,aps,pacs,twocolumn]{revtex4}
\usepackage{epsfig}
\usepackage{verbatim}
\usepackage{bm}
\usepackage{color}
\newcommand{\B}[1]{{\bm{#1}}}

\usepackage[latin1]{inputenc}
\newcommand{\beq}{\begin{equation}}
\newcommand{\eeq}{\end{equation}}
\newcommand{\bea}{\begin{eqnarray}}
\newcommand{\eea}{\end{eqnarray}}

\def\endproof{{\mbox{}\nolinebreak\hfill\rule{2mm}{2mm}\par\medbreak}}

\begin{document}
\title{Statistical Properties of Nonlinear Shell Models of Turbulence from Linear Advection Models: Rigorous Results}
\author{Roberto Benzi,$^1$, Boris Levant$^2$,  Itamar Procaccia$^3$
and Edriss S. Titi$^{2,4}$} \affiliation{$^1$ Dept. of Phyiscs, Universit\`a di Roma ``Tor Vergata"\\
$^2$ Dept. of Computer
Science and Applied Mathematics, $^3$Dept.  of Chemical Physics,
The Weizmann Institute of Science, Rehovot 76100, Israel,\\
$^4$Dept. of Mathematics and Dept. of Mechanical and Aerospace
Engineering, University of California, Irvine, CA 92697, USA}

\begin{abstract}
 In a recent paper it was proposed that for some {\bf nonlinear} shell models of turbulence
 one can construct a {\bf linear} advection model for an auxiliary field such that the scaling
 exponents of all the structure functions of the linear and nonlinear fields coincide.
 The argument depended on an assumption of continuity of the solutions as a function of a parameter. The aim of this paper is to provide a rigorous proof for the validity of the assumption. In
 addition
 we clarify here when the swap of a nonlinear model by a linear one will {\em not} work,

 \end{abstract}
\pacs{PACS number(s): 61.43.Hv, 05.45.Df, 05.70.Fh}
\maketitle

\section{Introduction}

Shell models of turbulence  \cite{GioBook,bif03} serve a  useful
purpose in studying the statistical properties of turbulent fields
due to their relative ease of simulation. In particular, shell
models allowed accurate direct numerical calculation of the scaling
exponents of their associated structure functions, including
convincing evidence for their universality
\cite{Gledzer,GOY,Jensen91PRA,Piss93PFA,Gat95PRE,sabra}. In
contrast, for the Navier-Stokes equations that model actual fluid
turbulence simulations are very much harder, and in addition one
still does not know whether these equations in 3-dimensions are
mathematically globally well posed. This problem does not exist in
shell models \cite{06CLT}, adding to their numerical attractiveness
a possibility to prove various properties and results rigorously
\cite{06CLT,06CLT_2}.

Consider for example the Sabra
shell model \cite{sabra} which, like other shell models of turbulence,
is a truncated description of the dynamics of Fourier modes,
preserving some of the structure and conservation laws of the
Navier-Stokes equations:
\begin{eqnarray}
(\frac{d }{dt}&+&\nu k_n^2 ) u_n = i\big[k_{n+1} u_{n+1}^* u_{n+2}-\delta
k_n u_{n-1}^* u_{n+1}\nonumber\\&+&(1-\delta) k_{n-1} u_{n-1} u_{n-2}\big]+f_n \ . \label{sabra}
\end{eqnarray}
Here $u_n$, with $n=0, 1, 2, \dots$ and the boundary conditions
$u_{-2} = u_{-1} = 0$, are the velocity modes restricted to
`wavevectors' $k_n=k_0 \mu^n$ with $k_0$ determined by the inverse
outer scale of turbulence. The model contains one additional
parameter, $\delta$, and it conserves two quadratic invariants (when
the force and the dissipation term are absent) for all values of
$\delta$. The first is the total energy $\sum_n |u_n|^2$ and the
second is $\sum_n (-1)^n k_n^{\alpha} |u_n|^2$, where $\alpha=
\log_{\mu} (1-\delta)$.

The scaling exponents are properties of the
structure functions. To define the structure function we introduce an average over time
according to
\begin{equation}
\langle A(t) \rangle  \equiv \lim_{T\to \infty} \case{1}{T}\int_0^T A(t) dt \ .
\end{equation}
In practice, however, we take
\begin{equation} \label{average}
\langle A(t) \rangle  = \case{1}{T}\int_0^T A(t) dt \ ,
\end{equation}
for some $T$ large enough, but finite. Our rigorous results also
refer to this definition of an average over time. For values of the
viscosity $\nu$ small enough, and for a sufficiently large amplitude
of the random force $f_n$ there exists a large range of values $k_n$
where numerical simulations show that structure functions follow a
power-law behavior. The low-order structure functions and the
associated scaling exponents are
\begin{eqnarray}
S_2(k_n)\equiv \langle u_n u^*_n\rangle \sim k_n^{-\zeta_2} \ , \label{S2}\\
S_3(k_n)\equiv \Im \langle u_{n-1}u_n u_{n+1}^*\rangle \sim k_n^{-\zeta_3}\ ,\label{S3}\\
\text{etc. for higher order}~S_p(k_n)\sim k_n^{-\zeta_p} \  . \nonumber
\end{eqnarray}
The values of the scaling exponents were determined accurately by
direct numerical simulations. Besides $\zeta_3$ which is exactly
unity \cite{Piss93PFA}, all the other exponents $\zeta_p$ appear anomalous, differing
from $p/3$. Numerical evidence is that the scaling exponents are also universal, i.e. they are
independent of the forcing $f_n$ as long as the latter is
restricted to small $n$ \cite{bif03}. It was shown that for $0<\delta<1$ the leading scaling exponents
are determined by the cascade of the energy invariant from large to small scales. For  $1<\delta<2$
the second invariant $\sum_n (-1)^n k_n^{\alpha} |u_n|^2$ becomes important in determining
the leading scaling exponents of the structure functions of $u_n$. In the bulk of this paper we
consider the situation with $0<\delta<1$, but return to the other case in Sect. \ref{2D}, in order
to clarify the role of invariants in determining the scaling properties.

In a recent publication further insight to the anomaly of the
exponents of the nonlinear problem (for the field $u_n$) was sought
by relating them to the scaling exponents of a {\em linear} model
for a field $w_n$. The linear model was constructed such that its
scaling exponents should be the same as those of the nonlinear
problem. The equations for this field are constructed under the
following requirements: (i) the structure of the equations is
obtained by linearizing the nonlinear problem and retaining only
such terms that conserve the energy; (ii) the resulting equation is
identical with the sabra model when $w_n = u_n$; (iii) the energy is
the only quadratic invariant for the passive field in the absence of
forcing and dissipation. These requirements lead to the following
linear model:
\begin{equation} \label{linear}
\frac{dw_n}{dt} = \frac{i}{3}\Phi_n(\B u,\B w)-\nu k_n^2 w_n +f_n\ , \label{lin}
\end{equation}
where the advection term is defined as
\begin{eqnarray}
&&\Phi_n(\B u,\B w) = k_{n+1}[(1+\delta) u_{n+2}w_{n+1}^*+(2-\delta)
u_{n+1}^*w_{n+2}]\nonumber\\
&&+ k_{n}[(1-2\delta)u_{n-1}^*w_{n+1}-(1+\delta)u_{n+1}w_{n-1}^*]\nonumber\\
&&+ k_{n-1}[(2-\delta)u_{n-1}w_{n-2}+(1-2\delta)u_{n-2}w_{n-1}] \ , \label{bilinear}
\end{eqnarray}
together with $u_{-1}=u_{-2}=w_{-1}=w_{-2}=0$. Observe that when
$w_n = u_n$ this model reproduces the Sabra model, and also that the
total energy is conserved because $\sum_n \Im[\Phi_n(u,w)w^*_n] =
0$. The second quadratic invariant is not conserved by the linear
model (\ref{linear}). Finally, both models have the same `phase
symmetry' in the sense that the phase transformations $u_n\to u_n
\exp{(i\phi_n)}$ and $w_n\to w_n\exp{(i\theta_n)}$ leave the
equations invariant iff $\phi_{n-1}+\phi_n = \phi_{n+1}$,
$\theta_{n-1}+\theta_n=\theta_{n+1}$. This identical phase
relationship
 guarantees that the non-vanishing correlation functions of both
models have precisely the same forms.
Thus for example the only  second and third
correlation functions in both models are those written explicitly in Eqs. (\ref{S2}) and (\ref{S3}).

The advantage of the linear model is that correlation functions are advanced
in time by a linear propagator \cite{01ABCPM,02CGP,fgv}. The linear model possesses ``Statistically
Preserved Structures" (SPS) which are evident in the decaying problem
Eq. (\ref{lin}) with $f_n=0$. These are {\em left} eigenfunctions of
eigenvalue 1 of the linear propagators for each order (decaying)
correlation function \cite{fgv}. For example for the second order
correlation function denote the propagator $P^{(2)}_{n,n'}(t|t_0)$;
this operator propagates any initial condition $\langle w_n
w^*_n\rangle (t_0)$ (with average over initial conditions, independent
of the realizations of the advecting field $u_n$) to the decaying
correlation function (with average over realizations of the advecting
field $u_n$)
\begin{equation}
\langle w_n w^*_n\rangle(t) = P^{(2)}_{n,n'}(t|t_0)\langle w_{n' }w^*_{n'}\rangle (t_0)  \ .
\end{equation}
The second order SPS, $Z^{(2)}_n$, is the left eigenfunction with eigenvalue 1,
\begin{equation}
Z^{(2)}_{n'} = Z^{(2)}_n P^{(2)}_{n,n'}(t|t_0) \ .
\end{equation}
Note that $Z^{(2)}_n$ is time independent even though the operator
$P^{(2)}_{n,n'}(t|t_0)$ is time dependent. Each order correlation
function is associated with another propagator $\B P^{(p)}(t|t_0)$ and
each of those has an SPS, i.e. a {\em left} eigenfunction $\B Z^{(p)}$
of eigenvalue 1. These non-decaying eigenfunctions scale with $k_n$,
$\B Z^{(p)}\sim k_n^{-\xi_p}$, and the values of the exponents $\xi_p$
are anomalous. Finally,
it was shown that  these SPS  are also the leading scaling
contributions to the structure functions of the {\em forced} problem (\ref{lin})  \cite{fgv,01ABCPM}.   Thus {\bf the scaling exponents of the linear problem are independent
of the forcing $f_n$}, since they are determined by the SPS of the
decaying problem.

\begin{figure}
\centering
\epsfig{width=.40\textwidth,file=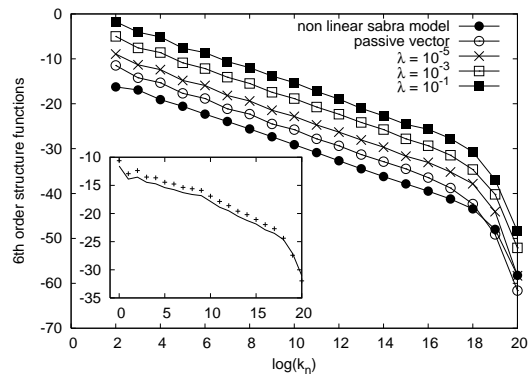}
\caption{ The sixth
order structure function of the field $w_n$ in Eqs. (\ref{system}) for
$\lambda=10^{-1}, 10^{-3}$ and $10^{-5}$,
together with the sixth order structure function for the
Sabra model (\ref{sabra})  and
for the linear model (\ref{lin}), respectively.
The structure function of the field $u_n$ for $\lambda>0$ are not shown since they are indistinguishable from those of the $w_n$.
Inset:  log-log plot of the fourth-order correlation function
$F_{2,2}(k_n,k_7)$ vs. $k_n$ calculated for the linear field ($+$) and for
the nonlinear field (solid line)  at $\lambda=0$.}
\label{lambdato0}
\end{figure}

To connect the linear model to the nonlinear problem one considers
the system of two coupled equations
\begin{eqnarray}
\frac{du_n}{dt} &=& \frac{i}{3}\Phi_n(u,u)+\frac{i \lambda}{3}\Phi_n(w,u)-\nu k_n^2 u_n +f_n\ , \nonumber\\
\frac{dw_n}{dt} &=& \frac{i}{3}\Phi_n(u,w)+\frac{i \lambda}{3}\Phi_n(w,w)-\nu k_n^2 w_n +\tilde f_n \label{system}
\end{eqnarray}
with $\lambda$ being a real parameter and $f_n$ and $\tilde f_n$ being different realizations
of the same random force. Observe that for any
$\lambda\ne 0$, the two equations in (\ref{system})
 exchange roles under the change $ \lambda w_n \leftrightarrow u_n$.
Thus if the scaling exponents $\xi_p$ and $\zeta_p$ of the two fields exist (i.e. a true scaling range exists), they must be the same for all $\lambda\ne 0$. For $\lambda=0$ we recover
the equations for the nonlinear and the linear models,
Eqs. (\ref{sabra}) and (\ref{lin}). In \cite{linnonlin} it was {\it assumed} that
the scaling exponents of either
field exhibits no jump in the limit $\lambda\to 0$. Indeed, Ref. \cite{linnonlin} also presented strong evidence for the validity of this assumption (see Fig. \ref{lambdato0}), but no mathematical proof was provided.

The aim of this paper is to close this gap. The main result of our
paper states that the solutions of the system (\ref{system}) exist
globally in time and depend continuously on the parameter $\lambda$,
including the limit $\lambda \to 0$. In particular we will show that
{\em if} the structure function of $u_n$ and $w_n$ exhibit the same
scaling exponents for any $\lambda\ne 0$, they will also have the
same scaling exponents  in the limit $\lambda\to 0$ (including
$\lambda=0$). We would like to stress here that our rigorous results
are valid for the structure functions, calculated over large, but
finite, fixed interval of time, which is consistent with the
definition (\ref{average}) of the long time average. In addition, we
would like to note that the numerical results correspond to the
equations with the stochastic implementation of the forcing, while
our rigorous proofs deal with the deterministic force that depend on
time. The statement and the proof of the main theorem will be given
in Section~\ref{section2}. The proof is based on the results on the
global existence and uniqueness of solutions of Eq. (\ref{sabra}),
obtained previously in~\cite{06CLT}. In the following section we
present the necessary mathematical definitions and formulate the
essential statement from~\cite{06CLT}.

\section{Functional Setting  and Previous Analytic Results}
\label{section1}

\subsection{Functional Setting}
Following the classical treatment of the Navier-Stokes and Euler equations we
re-write the sabra shell model for an infinite vector $\B u \equiv (u_0, u_1, \dots)$
\begin{equation}
\frac{d\B u}{dt}  + \nu \B A \B u + \frac{i}{3}\B \Phi(\B u,\B u) = \B f \ ,
\label{sabra2}
\end{equation}
together with the initial conditions $\B u(t=0)$. Introduce a Hilbert space $H$ which is the space of infinite vectors equipped with the scalar product $(\cdot, \cdot)$ and
the corresponding norm $|{\cdot}|$ defined as
\begin{equation}
(\B u, \B v) = \sum_{n=0}^\infty u_n v_n^*, \quad |\B u|= \bigg(
\sum_{n=0}^\infty |u_n|^2 \bigg)^{1/2},
\end{equation}
for every $\B u, \B v\in H$. The space $H$ is a space of all the velocity
vectors having finite energy.

The linear operator $\B A$, with a domain $D(\B A)$ dense in $H$, is
a positive definite, diagonal operator defined through its action on
the elements $\B u $ by
\begin{equation}
\B A \B u= (k_0^2 u_0,k_1^2 u_1, \dots)
\end{equation}
where the eigenvalues $k_n^2$ satisfy $k_n=k_0\mu^n$. Using the fact
that $\B A$ is a positive definite operator, we can define the
powers $\B A^s$ of $\B A$ for every $s\in {\mathcal R}$
\begin{equation}
\forall \B u = (u_0, u_1, u_2, \dots), \quad \B A^s u = (k_0^{2s} u_0,
k_1^{2s} u_1, k_2^{2s} u_2, \dots).
\end{equation}
The space $D(\B A^{s/2})$  is the domain of the operator $\B
A^{s/2}$ and we denote
\begin{equation}
V_s \equiv D(\B A^{s/2}) = \{ \B u = (u_0, u_1, u_2, \dots) \ ,
\sum_{j=0}^\infty k_j^{2s} |u_j|^2 < \infty \},
\end{equation}
which are Hilbert spaces equipped with the scalar product
\begin{equation}
(\B u, \B v)_{s} = (\B A^{s/2} \B u, \B A^{s/2} \B v), \;\;\; \forall \B u, \B v\in
D(\B A^{s/2}),
\end{equation}
and the norm $|\B u|_s^2 = (\B u, \B u)_s$, for every $\B u\in D(\B
A^{s/2})$. Since $V_{s} $ contains velocity vectors with $s$
``derivatives'',
\begin{equation}
V_{s} \subseteq V_0 = H \subseteq V_{-s}, \;\;\; \forall s > 0.
\end{equation}
The case of $s = 1$ is of a special interest for us. We denote $V =
D(\B A^{1/2})$ a Hilbert space equipped with a scalar product and a
corresponding norm
\begin{equation}
((\B u, \B v)) = (\B A^{1/2}\B u, \B A^{1/2}\B v), \;\;\; ||\B u||^2 = ((\B u, \B u)),
\end{equation}
for every $\B u, \B v\in V$. In addition the following relation
holds
\begin{equation} \label{norms}
k_1 |\B u| \le ||\B u|| \ .
\end{equation}
In what follows we will need the interpolation inequality for the
spaces $V_s$.

{\bf Lemma 1}. Let $s > 0$, then for all $\B u\in V_s$ and $0 < s' < s$
\begin{equation}
|\B u|_{s'} \le |\B u|^{1-s'/s} |\B u|_s^{s'/s} \ . \label{inter}
\end{equation}

{\it Proof}.  The Lemma follows by a simple application of H\"{o}lder inequality.
\endproof

The bilinear operator $\frac{i}{3}\B \Phi(\B u,\B w)$ is defined above, cf. Eq. (\ref{bilinear})
together with $\B\Phi\equiv (\Phi_0, \Phi_1, \dots)$. In \cite{06CLT} it was shown
that such a definition of the bilinear operator makes $\B\Phi$ and element of $H$ whenever
$\B u \in H$ and $\B w\in V$. For any $\B u,\B v \in H$ and $\B w \in V$ one proves  \cite{06CLT}
\begin{equation}
|(\B \Phi(\B u, \B v), \B w)| \le C |\B u|~|\B v|~ ||\B w|| \ . \label{ineq}
\end{equation}
for some positive constant $C$. In addition the conservation law is
written is the present notation as
\begin{equation}
\Im (\B \Phi(\B u, \B w), \B w)=0 \ . \label{conserve}
\end{equation}

All our rigorous results are valid for the deterministic forcing $\B
f$. Therefore, in order to account for the stochastic implementation
of the forcing term we will assume that $\B f$ depends on time, but
always stays bounded in the $H$-norm. More precisely, define the
space $L^\infty([0, T), H)$, for some $0 < T \le \infty$, as a space
of functions of the time interval $[0, T)$ with values in $H$ and
the norm defined as
\begin{equation} \label{inftynorm}
|| \B f ||_\infty = \sup_{0 \le t < T} | \B f(t) | \ .
\end{equation}
In what follows, we will assume that the forcing term $\B f$
satisfies $\B f\in L^\infty([0, \infty), H)$.

\subsection{Summary of Previous Results}

In \cite{06CLT} Eq. (\ref{sabra}) was studied, and the relevant results can be formulated
as the following theorem:

{\bf Theorem (CLT 06)}: The solution of Eq. (\ref{sabra}) exists
globally in time for any initial condition $\B u(0)$ in H, and for
any $\B f$ in $L^\infty([0, \infty), H)$. Moreover, the solutions
are unique and the energy of the solution $\B u(t)$ is bounded for
all times:
\begin{equation} \label{bound}
\sup_{0 \le t \le T} |\B u(t)|^2 \le K_0(k_0,\mu,\nu,\delta, \B
u(0), \B f) \ ,
\end{equation}
where the \textit{a-priori} constant $K_0$ depends on all the
parameters of the equations, on the forcing and on the initial
conditions.

If in addition we assume that the forcing $\B f=(f_0, f_1, \dots)$
acts on the finite number of shells, namely, if there exists $N \ge
0$, such that $f_n = 0$, for all $n\ge N$, then the solution $\B
u=(u_0,u_1, \dots)$ has an exponentially decaying spectrum $|u_n|$
as a function of $k_n$, and in particular
\begin{equation}
\sup_{\nu k_0^2 \le t \le T} |\B u(t)|_s^2 \le
K_s(k_0,\mu,\nu,\delta, \B u(0), \B f),
\end{equation}
for any $s > 0$ and the \textit{a-priori} constants $K_s$ depend on
all the parameters of the equations, on the forcing and on the
$L^\infty([0, \infty), H)$-norm of the initial conditions (see
definition (\ref{inftynorm})).

\section{The Main Result} \label{section2}
The main statement of this paper is that the system of equations
(\ref{system}) is globally well posed for all real $\lambda$, and
that the solutions depend continuously on the parameter $\lambda$.
In particular, as $\lambda\to 0$ the solution of the system converge
uniformly on any finite interval of time to the solution of the
system with $\lambda=0$. This statement is formulated as follows:

{\bf Theorem 1}. Let $\B u(0;\lambda), \B w(0;\lambda)$ be given in
$H$, and the forces $\B f, \B {\tilde f}$ in $L^\infty([0, T], H)$.
Denote by $\B u(t;\lambda), \B w(t;\lambda)$ the solutions of the
coupled system (\ref{system}). Then the following holds

\begin{enumerate}
\item \label{thm_main_2} The solution of the system (\ref{system}) exists globally in
time for every $\lambda$. Moreover, energy of the solutions $|\B
u(t; \lambda)|^2$ and $|\B w(t; \lambda)|^2$ are bounded for all
times, where the bounds depend on all the parameters of the
equations, $\lambda$, the forcing and the initial conditions.

\item \label{thm_main_4} For any $T > 0$,
\[
\lim_{\lambda \to 0} \sup_{0 \le t \le T} |\B u(t;\lambda) - \B u(t;
0)|^2 = 0,
\]
and
\[
\lim_{\lambda \to 0} \sup_{0 \le t \le T} |\B w(t;\lambda) - \B
w(t;0)|^2 = 0,
\]
where $\B u(t; 0), w(t; 0)$ are the solutions of (\ref{system}) with
$\lambda = 0$.

\item \label{thm_main_3} If in addition we assume that the forces
$f, {\tilde f}$ act only on the finite number of shells then for any
$T \ge \nu k_0^2$, and $s > 0$,
\[
\lim_{\lambda \to 0} \sup_{\nu k_0^2 \le t \le T} |\B u(t;\lambda) -
\B u(t; 0)|_s^2 = 0,
\]
and
\[
\lim_{\lambda \to 0} \sup_{\nu k_0^2 \le t \le T} |\B w(t;\lambda) -
\B w(t;0)|_s^2 = 0,
\]
where $\B u(t; 0), w(t; 0)$ are the solutions of (\ref{system}) with
$\lambda = 0$.
\end{enumerate}

{\it Proof}: Part 1. of this theorem follows from defining a new
variable $\B q(t;\lambda)\equiv \B u (t;\lambda) +\lambda \B w
(t;\lambda)$. This variable satisfies Eq. (\ref{sabra}) with a
forcing $\B f +\lambda \B {\tilde f}$. Accordingly theorem CLT06
provides the proof that $\B q(t; \lambda)$ exists globally in time
for every $\lambda$. Next observe that the system of equations
(\ref{system}) can be rewritten in the form
\begin{eqnarray}
\frac{du_n}{dt} &=& \frac{i}{3}\Phi_n(\B q,\B u))-\nu k_n^2 u_n +f_n\ , \nonumber\\
\frac{dw_n}{dt} &=& \frac{i}{3}\Phi_n(\B q,\B w)-\nu k_n^2 w_n
+\tilde f_n \ . \label{system1}
\end{eqnarray}
This form of writing shows that the the fields $\B u$ and $\B w$
satisfy {\em linear} diffusion advection equations advected by a
smooth field $\B q$. Accordingly both fields remain smooth for all
time and all $\lambda$. In addition, using the relation
(\ref{bound}) one is able to derive the bounds for the energies of
the solutions.

To prove Part 2. of the theorem we need first \\
{\bf Proposition 1}: Denote by $\B q(t;0)$ the solution of the Sabra
model (\ref{sabra}) with initial data $\B q(0;0) $ in $H$. That  is
also the solution of the first of equations (\ref{system}) when
$\lambda = 0$. Then
\[
\lim_{\lambda\to 0} \sup_{0 \le t \le T} |\B q(t;\lambda) - \B
q(t;0)|= 0.
\]
Moreover, if $\B f,$ and $\tilde{\B f}$ act only on finite number of
shells, then for any $s > 0$ we have
\[
\lim_{\lambda\to 0} \sup_{\nu k_0^2 \le t \le T} |\B q(t;\lambda) -
\B q(t;0)|_s = 0.
\]

{\it Proof}: Let us denote by $\Delta \B q = \B q(t;\lambda) - \B
q(t;0)$. Then $\Delta \B q$ satisfies the equation
\begin{eqnarray}
&&\frac{d\Delta \B q }{dt}+ \nu\B A \Delta \B q  -  \frac{i}{3}\B \Phi(\B q(t;\lambda), \Delta \B q) -  \frac{i}{3}\B \Phi(\Delta \B q, \B q(t;\lambda)) \nonumber\\&&+  \frac{i}{3}\B \Phi(\Delta \B q, \Delta \B q) = \lambda\B{ \tilde f} \label{diff} \\
&&\Delta \B q(0) = \lambda \B w(0). \label{diff1}
\end{eqnarray}

Take now the inner product in $H$ of the above
equation with $\Delta \B q$. Computing the real part and using Eq. (\ref{conserve}) we find
\begin{equation}
\case{1}{2}\case{d}{dt}|\Delta \B q|^2 + \nu ||\Delta \B q||^2  -
\Re ( \frac{i}{3}\B \Phi(\Delta \B q, \B q(t;\lambda), \Delta \B q)
=\Re (\lambda\B{\tilde f},\Delta \B q)\ .
\end{equation}
Using the Cauchy-Schwarz inequality and the relation (\ref{ineq}) we
get
\begin{eqnarray}
&&\case{1}{2}\case{d}{dt}|\Delta \B q|^2 + \nu ||\Delta \B q||^2 \le
\nonumber\\
&&\le |(\frac{1}{3}\B \Phi(\Delta \B q, \B q(t;\lambda), \Delta \B
q)| + |(\lambda
\tilde f, \Delta \B q)| \nonumber\\
&&\le C |\Delta \B q|~||\Delta \B q||~ |q(t;\lambda)|+ \lambda |\B
{\tilde f}| ~|\Delta \B q| \ ,  \label{eq1}
\end{eqnarray}
Applying the Young's inequality ($ab \le a^2/2 + b^2/2$) twice and
using the inequality (\ref{norms}), we have
\begin{widetext}
\[
\case{1}{2}\case{d}{dt}|\Delta \B q|^2 + \nu ||\Delta \B q||^2 \le
\frac{C^2}{\nu} |\Delta \B q|^2 |q(t;\lambda)|^2 +
\frac{\lambda^2}{\nu k_1^2} |\B {\tilde f}|^2 + \frac{\nu}{2}
||\Delta \B q||^2.
\]
\end{widetext}
Using the fact that the relation (\ref{bound}) holds for any $t\ge
0$ and the fact that $\B f\in L^\infty([0, \infty), H)$ (see
definition (\ref{inftynorm})), we obtain
\[
\case{d}{dt}|\Delta \B q|^2 \le \frac{2 K_0^2 C^2}{\nu} |\Delta q|^2
+ \frac{2 \lambda^2}{\nu k_1^2} ||\B {\tilde f}||_\infty^2.
\]
By Gronwall's inequality we conclude
\begin{eqnarray}
&&|\Delta \B q(t)|^2  \le \nonumber\\
&&\le e^{\frac{2 K_0^2 C^2}{\nu} t} |\Delta \B q(0)|^2 +
\frac{\lambda^2 ||\B {\tilde f}||_\infty^2 }{K_0^2 C^2 k_1^2}
(e^{\frac{2 K_0^2 C^2}{\nu} t} - 1) = \nonumber
\\
&& = e^{\frac{2 K_0^2 C^2}{\nu} t} \lambda^2 |\B w(0)|^2 +
\frac{\lambda^2 ||\B {\tilde f}||_\infty^2 }{K_0^2 C^2 k_1^2}
(e^{\frac{2 K_0^2 C^2}{\nu} t} - 1).  \label{endprop}
\end{eqnarray}
Therefore, for any $T > 0$,
\[
\lim_{\lambda \to 0} \sup_{0 \le t \le T} |\Delta \B q(t)|^2 =
0,
\]
and the first statement of the proposition follows.

The second statement of the proposition follows from the boundedness
of $\B q(t;\lambda)$ and $\B q(t;0)$ in higher order norms after a
short transient period of times, say $\nu k_0^2$, (provided that the
forces $\B f,$ and $\tilde{\B f}$ act on the finite number of shells
as required by Theorem CLT06) and the interpolation inequality
(\ref{inter}).
\endproof

Finally, we are ready to finish the proof of the main theorem. Let
us fix $s \ge 0$, and show that
\begin{equation}
\lim_{\lambda \to 0} \sup_{\nu k_0^2 \le t \le T} |\B w(t;\lambda) -
\B w(t;0)|_s^2 = 0.
\end{equation}
Denote $\Delta \B w = \B w(t;\lambda) - \B w(t;0)$. The function
$\Delta \B w$ satisfies the equation
\begin{eqnarray}
&&\case{d}{dt} \Delta \B w + \nu \B A \Delta \B w - \frac{i}{3}\B
\Phi\big(\B q(t;\lambda),\B w(t;\lambda)\big)
\nonumber\\
&&+\frac{i}{3}\B \Phi\big(\B q(t;0),\B  w(t;0)\big) = 0.
\end{eqnarray}
We rewrite it in the form
\begin{widetext}
\begin{equation}
\case{d}{dt}  \Delta \B w + \nu \B A \Delta \B w -  \frac{i}{3}\B
\Phi\big(\B q(t;\lambda) - \B q(t;0), \B w(t;0)\big) - \frac{i}{3}\B
\Phi\big(\B q(t;\lambda), \Delta \B w\big ) = 0,
\end{equation}
and as before, taking the inner product in $H$ with $\Delta \B w$,
computing the real part and using Eq. (\ref{conserve}), we obtain
\[
\case{1}{2}\case{d}{dt}|\Delta \B w|^2 + \nu ||\Delta \B w||^2 -
\Re\Big( \frac{i}{3}\B \Phi\big(\B q(t;\lambda) - \B q(t;0), \B
w(t;0)\big), \Delta \B w\Big) = 0.
\]
Applying, subsequently the Young's inequality and the inequality
(\ref{ineq}), we get
\begin{eqnarray}
&&\case{1}{2}\case{d}{dt}|\Delta \B w|^2 + \nu ||\Delta \B w||^2 \le
|\big( \frac{i}{3}\B \Phi\big(\B q(t;\lambda) - \B q(t;0), \B
w(t;0)\big), \Delta \B w \big)| \le \nonumber \\ &&\le C ||\Delta \B
w||~ |\B w(t;0)| ~|\B q(t;\lambda) - \B q(t;0)| \le \frac{C^2}{2
\nu} |\B w(t;0)|^2 ~|\B q(t;\lambda) - \B q(t;0)|^2 + \frac{\nu}{2}
||\Delta\B  w||^2.
\end{eqnarray}
\end{widetext}
It follows that
\[
\case{d}{dt}|\Delta \B w|^2\le \frac{C^2}{\nu} |\B w(t;0)|^2 |\B
q(t;\lambda) - \B q(t;0)|^2,
\]
and integrating over $(0, t)$ we conclude
\[
|\Delta \B w(t)|^2 \le \frac{C^2 M}{\nu} t |\B q(t;\lambda) -\B
q(t;0)|^2,
\]
where we used the fact that $\Delta \B w(0) = 0$ and $|\B w(t;0)|^2
\le M$, for some constant $M$, as was stated in the Part~1 of the
main theorem. It follows from Proposition~1 that $\sup_{0 \le t \le
T} |\B q(t;\lambda) - q(t;0)|^2 \to 0$, as $\lambda\to 0$. Hence we
may conclude that
\[
\lim_{\lambda \to 0} \sup_{0 \le t \le T} |\Delta \B w(t)|^2 =
0.
\]
By virtue of the interpolation inequality (\ref{inter}) one can
follow steps as in the proof of Proposition $1$ to show that the
higher order norms of $\Delta \B w$ also converge to $0$ uniformly
in the time interval $\nu k_0^2 \le t \le T$, as $\lambda \to 0$. To
finish the proof, on can observe that
\begin{eqnarray}
&&|\B u(t;\lambda) - \B u(t;0)| = |\B q(t;\lambda) - \B q(t;0) -
\lambda \B w(t;\lambda)|
 \nonumber\\&&\le |\B q(t;\lambda) - \B q(t;0)|+ \lambda |\B w(t;\lambda)|,
\end{eqnarray}
and from the fact that $|\B w(t;\lambda)|$ is globally bounded in
time it follows from Proposition~1 that
\[
\lim_{\lambda \to 0} \sup_{0 \le t \le T} |\B u(t;\lambda) - \B u(t;0)| =
0.
\]
\endproof
\begin{figure}
\centering
\epsfig{width=.40\textwidth,file=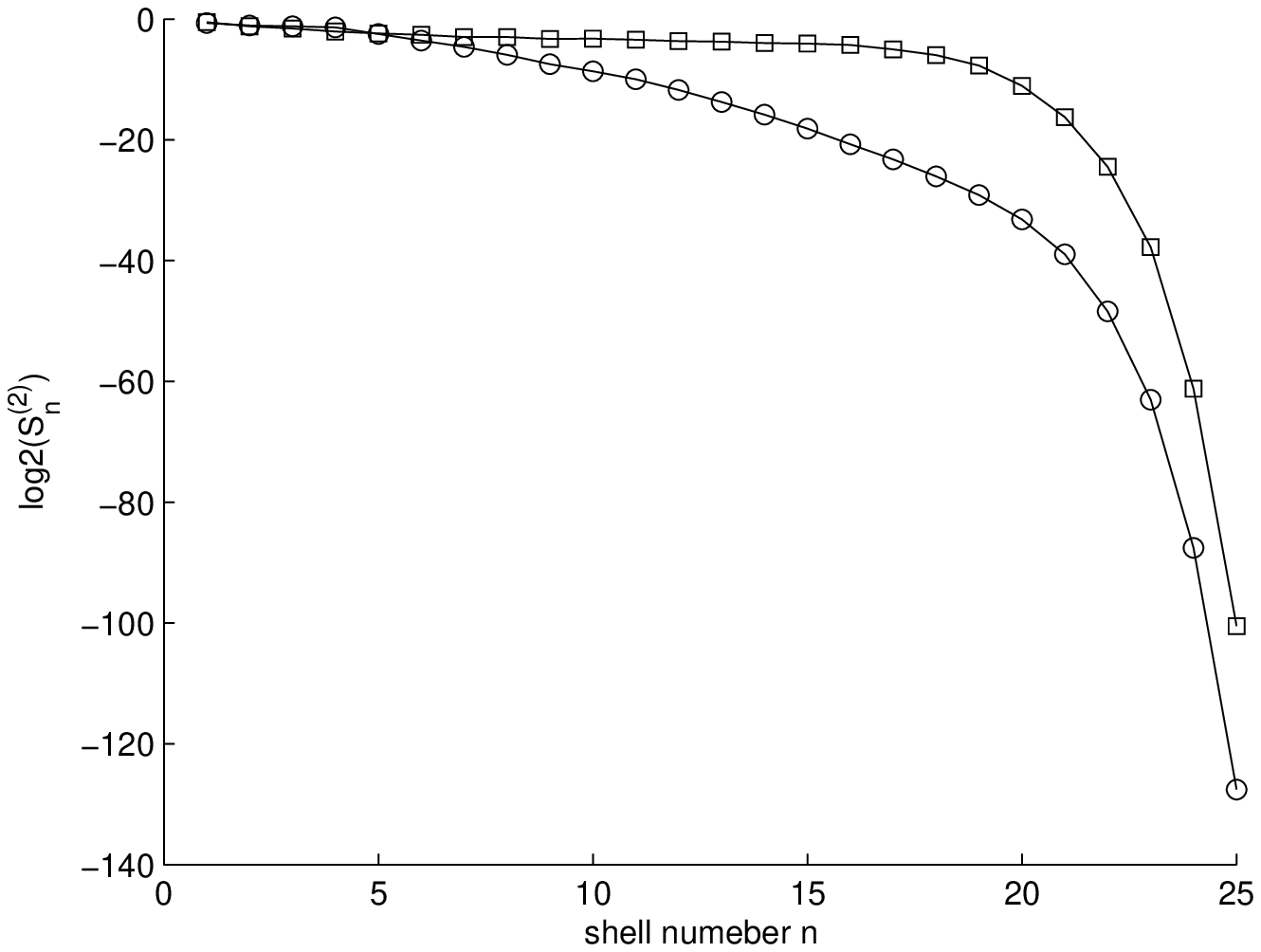}
\caption{ Double logarithmic plot of the second order structure functions $\langle u_n u^*_n\rangle$ and
$\langle w_n w^*_n\rangle$ of the system of equations (\ref{system}) for $\delta=1.25$ and
$\lambda=0$.}
\label{2d-lam=0}
\end{figure}
\begin{figure}
\centering
\epsfig{width=.40\textwidth,file=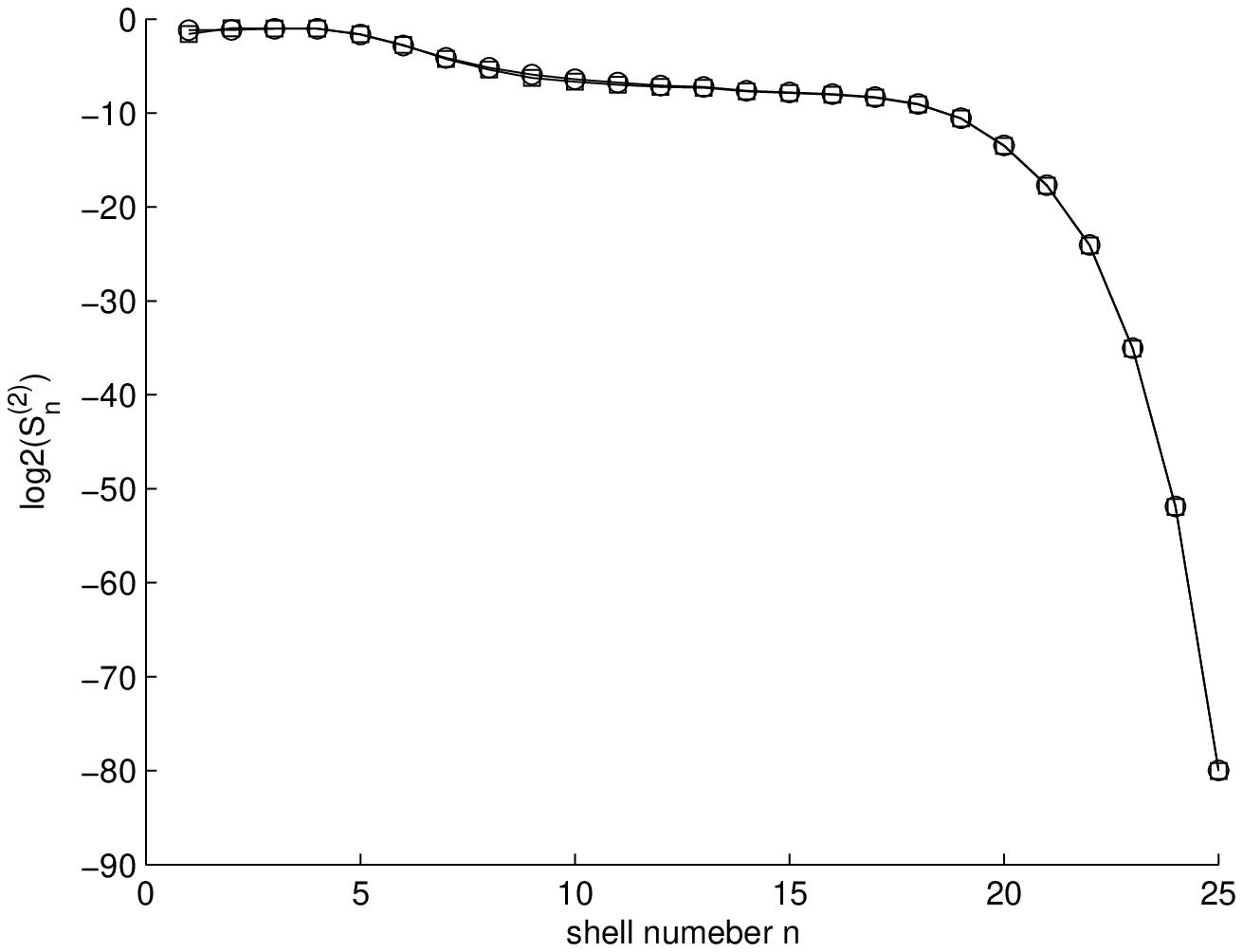}
\caption{ Double logarithmic plot of the second order structure functions $\langle u_n u^*_n\rangle$ and
$\langle w_n w^*_n\rangle$ of the system of equations (\ref{system}) for $\delta=1.25$ and
$\lambda=1$.}
\label{2d-lam=1}
\end{figure}
\begin{figure}
\centering
\epsfig{width=.40\textwidth,file=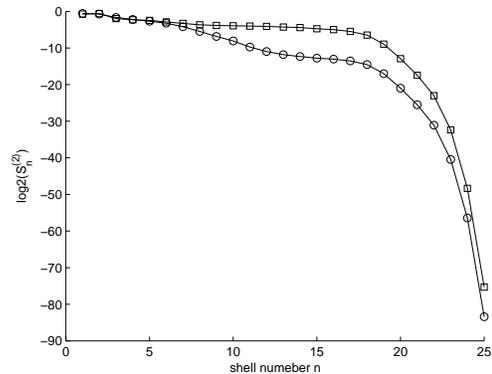}
\epsfig{width=.40\textwidth,file=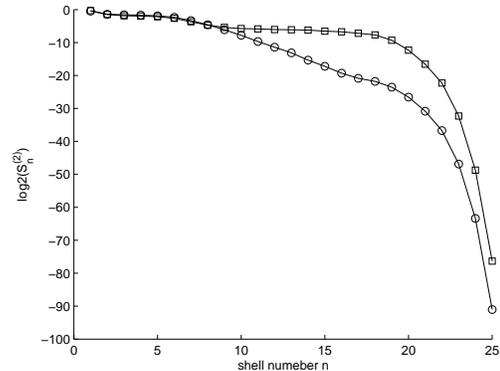}
\caption{ Double logarithmic plot of the second order structure functions $\langle u_n u^*_n\rangle$ and
$\langle w_n w^*_n\rangle$ of the system of equations (\ref{system}) for $\delta=1.25$ and
$\lambda=0.1$ (upper panel) and $\lambda=0.01$ (lower panel).}
\label{2d-lamsmall}
\end{figure}
\subsection{Consequences for Structure Functions}

Using the form of Eqs. (\ref{system1}) we conclude that for $f_n$
and $\tilde f_n$ being different realizations of the same random
force, whenever the structure functions of $u_n$ and $w_n$ exhibit
the same scaling exponents for all finite $\lambda$, they must also
exhibit the same scaling exponents for $\lambda=0$.

\section{When can the Nonlinear Model exhibit Scaling Exponents that are Different from
the Linear Model}
\label{2D}

In this section we turn our attention to situations when the nonlinear and the linear models
cannot exhibit the same scaling exponents. The theorem as stated and proven still holds, but as we shall see this is a situation for which the two fields $u_n$ and $w_n$ cannot exhibit the same scaling
exponents for all $\lambda\ne 0$. A case in point is the same nonlinear Sabra model with $1<\delta<2$.
In Fig. \ref{2d-lam=0} we show the second order structure functions $\langle u_n u^*_n\rangle$ and
$\langle w_n w^*_n\rangle$ obtained from simulating Eqs. (\ref{system}) for $\lambda=0$ and
$\delta=1.25$.  As expected, the scaling of $w_n$ is influenced by the cascade of energy, whereas
that of $u_n$ by the cascade of the second invariant. As a result the scaling exponents are
distinctly different. The same system of equations for $\lambda=1$ is symmetric in $w_n$ and
$u_n$. Indeed, in Fig. \ref{2d-lam=1} we show the result of simulations for $\lambda=1$, for which the second order structure function of the two fields is identical. Now however we cannot expect that
this identity will persist for $\lambda\to 0$. In Fig, \ref{2d-lamsmall} we show the results of simulations
for the same system of equations for $\lambda=0.1$ and $\lambda=0.01$.

To understand the results of the simulations we note that when $\lambda\ne 0$ the second
invariant of the equation for $u_n$ is destroyed, and one could think that the scaling
of $u_n$ should be dominated by the energy invariant. This is certainly true for $\lambda=1$.
But now when $\lambda$ decreases towards zero, we should reconsider the system of equations
(\ref{system}) by renaming $\tilde w_n=\lambda w_n$. Substituting this re-named variables into
Eqs. (\ref{system}) and re-arranging, we read
\begin{eqnarray}
\frac{du_n}{dt} &=& \frac{i}{3}\Phi_n(u,u)+\frac{i }{3}\Phi_n(\tilde w,u)-\nu k_n^2 u_n +f_n\ , \nonumber\\
\frac{d\tilde w_n}{dt} &=& \frac{i}{3}\Phi_n(u,\tilde w)+\frac{i
}{3}\Phi_n(\tilde w,\tilde w)-\nu k_n^2 \tilde w_n +\lambda \tilde
f_n  \ . \label{system2}
\end{eqnarray}
Thus the net result of the transformation is that the equations for $u_n$ and $\tilde w_n$ are the same,
but the forcing of $w_n$ becomes weaker and weaker as $\lambda\to 0$. Accordingly, while the
second invariant is still destroyed as a true invariant for any value of $\lambda$, for small $\lambda$
the strength of the term $\frac{i }{3}\Phi_n(\tilde w,u)$ diminishes, allowing a cross-over behavior
in the scaling of $u_n$, precisely as we see in Fig. \ref{2d-lamsmall}.

\acknowledgments

IP acknowledges useful discussions with L. Angheluta, especially
concerning the material presented in Sec. IV. This work has been
supported in part by the US-Israel Binational Science Foundation.
The work of E.S.T. was also supported in part by the NSF grant
No.~DMS--0504619, and by the ISF grant No.~120/06.


\begin{thebibliography}{99}

\bibitem{GioBook}T. Bohr, M.H.  Jensen, G.  Paladin, A.  Vulpiani,
{\it Dynamical systems approach to turbulence} (Cambridge University
Press, 1998)
\bibitem{bif03} L. Biferale
Ann. Rev. Fluid. Mech.  {\bf 35}, 441,  (2003).

\bibitem{Gledzer}E.~B. Gledzer.
Dokl. Akad. Nauk. SSSR, {\bf 20}, 1046 (1973).

\bibitem{GOY}
M.~Yamada and K.~Ohkitani.
J. Phys. Soc. Jpn {\bf 56}, 4210 (1987).

\bibitem{Jensen91PRA}
M.~H. Jensen, G.~Paladin, and A.~Vulpiani.
Phys. Rev. A, {\bf 43}, 798 (1991).

\bibitem{Piss93PFA}
D.~Pissarenko, L.~Biferale, D.~Courvoisier, U.~Frisch, and M.~Vergassola.
 Phys. Fluids A {\bf 5}, 2533 (1993).

\bibitem{Gat95PRE}
I.~Procaccia O.~Gat and R.~Zeitak.
Phys. Rev. E ,  {\bf 51}, 1148 (1995).

\bibitem{sabra}
 V.S. L'vov, E. Podivilov, A. Pomyalov, I. Procaccia and D. Vandembroucq,  Phys. Rev. E , {\bf 58}  1811(1998).

 \bibitem{Benzi93PHD}
R.~Benzi, L.~Biferale, and G.~Parisi.
Physica D {\bf 65}, 163 (1993).

\bibitem{Benzi1}
R. Benzi, L. Biferale  and F. Toschi
Eur. Phys. J. B  {\bf 24},   125,  (2001).

\bibitem{Benzi2}R. Benzi, L. Biferale, M. Sbragaglia and F. Toschi
Phys. Rev. E {\bf 68},  046304, (2004).


\bibitem{ditlev} P. Ditlevsen, Phys. Rev. E, {\bf 54} 985, (1996).

\bibitem{pierotti}
L. Biferale, D. Pierotti and F. Toschi
Phys. Rev. E  {\bf 57},   R2515, (1998).


\bibitem{fgv} G. Falkovich, K. Gawedzk, M. Vergassola,
  Rev. Mod.  Phys. {\bf 73} 913 (2001).

 \bibitem{01ABCPM}
 I. Arad, L. Biferale, A. Celani, I. Procaccia, and M. Vergassola, Phys. Rev. Lett., {\bf 87}, 164502 (2001) .

 \bibitem{02CGP}
 Y. Cohen, T. Gilbert and I. Procaccia, Phys. Rev. E., {\bf 65},026314 (2002).

  \bibitem{01AP}
 I. Arad and I. Procaccia, Phys. Rev. E, {\bf 63}, 056302 (2001).

\bibitem{CV} A. Celani and M. Vergassola, Phys. Rev. Lett., {\bf 86 } 424
(2001).

 \bibitem{03CPP}
 Y. Cohen, A. Pomyalov and I. Procaccia,  Phys. Rev. E., {\bf 68}, 036303 (2003).

 \bibitem{linnonlin}
 L. Angheluta, R. Benzi, L. Biferale, I. Procaccia and F. Toschi, Phys. Rev. Lett., in press.

 \bibitem{06CLT}
 P. Constantin, B. Levant and E. S. Titi, Physica D {\bf 219} (2006), 120-141.

 \bibitem{06CLT_2}
 P. Constantin, B. Levant and E. S. Titi, Phys. Rev. E, accepted.

\end{thebibliography}
\end{document}